\documentclass[lettersize,journal]{IEEEtran}
\usepackage{amsmath,amsfonts}
\usepackage{amssymb}
\usepackage{algorithmic}
\usepackage{array}
\usepackage[caption=false,font=normalsize,labelfont=sf,textfont=sf]{subfig}
\usepackage{booktabs}
\usepackage{textcomp}
\usepackage{makecell}
\usepackage{stfloats}
\usepackage{url}
\usepackage{xcolor}
\usepackage[table]{xcolor}
\usepackage{pgf}
\usepackage{colortbl}
\usepackage{verbatim}
\usepackage{graphicx}
\usepackage{multirow}
\usepackage{ragged2e}
\def\BibTeX{{\rm B\kern-.05em{\sc i\kern-.025em b}\kern-.08em
    T\kern-.1667em\lower.7ex\hbox{E}\kern-.125emX}}
\newcommand{\smallpercent}[1]{{\scriptsize #1}}
\newcommand{\diffcell}[2]{%
	\pgfmathparse{min(10 + (#1 / 20.0) * 90.0, 100)}%
	\let\mycolorintensity\pgfmathresult
	\edef\tempcellcmd{\noexpand\cellcolor{green!\mycolorintensity!white}}%
	\tempcellcmd
	#2 \smallpercent{(#1\%)}%
}

\makeatletter
\let\oldbibitem\bibitem

\renewcommand{\bibitem}[2][]{%
	\normalcolor 
	
	\@ifundefined{bibcolor@#2}{}{%
		\color{\csname bibcolor@#2\endcsname}%
	}%
	
	\if\relax\detokenize{#1}\relax
	\oldbibitem{#2}%
	\else
	\oldbibitem[#1]{#2}%
	\fi
}

\newcommand{\colorbib}[2]{%
	\expandafter\def\csname bibcolor@#1\endcsname{#2}%
}
\makeatother
\usepackage{balance}
\begin{document}
\title{Adaptive Spatial-Temporal Graph Learning-Enabled Short-Term Voltage Stability Assessment against Time-Varying Topological Conditions}
\author{Chao Deng, \emph{Student Member, IEEE}, Lipeng Zhu, \emph{Senior Member, IEEE}, Chang Liu, Hefeng Zhai, Baoye Tian, Zexiang Zhu, Jiayong Li, \emph{Member, IEEE}, and Cong Zhang, \emph{Member, IEEE}
\thanks{
This work was supported by State Key Laboratory of HVDC under Grant SKLHVDC-2024-KF-04. (\emph{Corresponding author: Lipeng Zhu})

Chao Deng, Lipeng Zhu, Chang Liu, Jiayong Li, and Cong Zhang are with the College of Electrical and Information Engineering, Hunan University, Changsha 410082, China (e-mail: dengchaoll@hnu.edu.cn; lpzhu@hnu.edu.cn; 2024lc@hnu.edu.cn; j-y.li@connect.polyu.hk; zcong@hnu.edu.cn). 

Hefeng Zhai, Baoye Tian, and Zexiang Zhu are with State Key Laboratory of HVDC (China Southern Power Grid Electric Power Research Institute), Guangzhou 510663, China
(e-mail: zhaihf@csg.cn; tianby@csg.cn; zhuzx@csg.cn).}}

%\markboth{Journal of \LaTeX\ Class Files,~Vol.~18, No.~9, September~2020}%
%{How to Use the IEEEtran \LaTeX \ Templates}
\maketitle

\begin{abstract}
The emerging deep learning (DL) technology has recently exhibited great potential in data-driven short-term voltage stability (SVS) assessment of complex power grids. However, without sufficient attention to the time-varying topological structures of today’s power grids, the majority of existing DL-based SVS assessment schemes could experience severe performance degradation in practice. To address this drawback, this paper proposes an adaptive spatial-temporal graph learning-enabled SVS assessment approach that can adapt well to various topological changes. First, considering the time-varying topological conditions of a given power grid, an adaptive graph representation matrix is automatically learned to effectively capture the complicated spatial correlations between individual buses within the grid. Then, to help better capture regional SVS features for subsequent learning processes, the adaptive graph representation matrix is properly adjusted by introducing a spatial attention mechanism. Further, with post-fault system trajectory data linked together via attention-based graph representation, a residual spatiotemporal graph convolutional network is carefully built with Optuna-based optimization to deeply mine system-wide spatiotemporal features and thus achieve structure-adaptive SVS assessment. Numerical test results on two representative sub-systems of a realistic provincial power grid in South China demonstrate the efficacy of the proposed approach under various changing topological conditions.
\end{abstract}

\begin{IEEEkeywords}
Adaptive spatial-temporal graph learning, graph representation matrix, short-term voltage stability, time-varying topological conditions, attention mechanism.
\end{IEEEkeywords}

\section{Introduction}
\IEEEPARstart{I}{N} recent decades, due to the continuous expansion of electric power grids’ scale, the growing complexity of network structures, and the high penetration of renewable energy, the operating conditions and scenarios of modern power systems have been becoming increasingly complex and variable \cite{ref1}. Consequently, the systems’ security and stability during daily operations are largely challenged by large-disturbance stability issues, e.g., the short-term voltage stability (SVS) problem \cite{ref2}. Generally, the SVS problem is closely related to dynamic components in the system with rapid responses, e.g., induction motor loads and high-voltage direct current (HVDC) converters. The fast-acting behaviors of these components in transient processes induce much higher reactive power demands than those under steady-state operating conditions \cite{ref3}. During online monitoring, if not properly monitored and controlled, the SVS problem could lead to voltage collapse and even severe blackouts. Thus, timely and reliable online SVS assessment is of vital importance to ensure the secure and stable operation of power systems.

For SVS assessment, with the rise of advanced machine learning (ML) technologies in recent decades, a considerable number of inspiring ML-based solutions have been reported \cite{ref4}, \cite{ref5}, \cite{ref6}, \cite{ref7}, \cite{ref8}, \cite{ref9}, \cite{ref10}, \cite{ref11}, \cite{ref12}, including decision tree (DT), random forest (RF), convolutional neural network (CNN), long short-term memory (LSTM) network, and Transformer. However, with feature learning and representation in regular Euclidean spaces (e.g., flattened vectors, matrix-like images, and temporal sequences), and these solutions may fail to capture the inherent non-Euclidean spatial correlations in practical power grids with irregular network structures.
	
To address this defect, graph neural networks (GNNs) specializing in learning from non-Euclidean structural data have been widely adopted, yielding promising results in SVS assessment \cite{ref13}, \cite{ref14}, \cite{ref15}, \cite{ref16} as well as transient stability assessment (TSA) \cite{ref17}, \cite{ref18}, \cite{ref19}, \cite{ref20}. However, most GNN-based methods rely on a pre-defined and fixed graph structure, making it difficult to adapt to time-varying topological conditions in practical power grids. In fact, recent statistics \cite{ref21}-\cite{ref22} indicate that power systems in China experience an average of more than 10 switching-related topological changes per day. These changes typically stem from scheduled switching operations by system operators, line outages due to unexpected faults, and line outages caused by extreme weather conditions. Similar data have been reported in France \cite{ref23}, where the grids are operated with $5\!\sim\!20$ topological switching actions per day to accommodate the increasing penetration of fluctuating renewable energy. If the changing characteristics of system topologies are not sufficiently considered during offline learning, the resulting SVS assessment models may undergo performance degradation under online varying topological conditions. Although some studies preliminarily analyze the impact of topological changes on online SVS assessment performance \cite{ref5}, \cite{ref13}, \cite{ref20}, they do not provide effective solutions to actively address the structural changing issue.

In recent years, a handful of research efforts have been made to realize structure-adaptive SVS assessment via transfer learning (TL) techniques \cite{ref24}, \cite{ref25}, \cite{ref26}, \cite{ref27}. However, TL schemes require generating new samples and repeatedly implementing retraining for new topological scenarios, which makes their reliability highly dependent on the knowledge of the accurate topological change information during online monitoring. In addition, the repeated model fine-tuning in TL-based schemes increases the computational burden and is difficult for online SVS assessment in complex systems with frequently changing topologies. Overall, how to construct a structure-adaptive SVS assessment scheme that does not rely on exact topological information and can automatically perceive system topological changes during online monitoring remains to be a practical yet unsolved challenge.

To tackle the above-mentioned research gap, this paper proposes an adaptive spatio-temporal graph learning (ASTGL)-enabled online SVS assessment approach. Unlike the aforementioned transfer learning schemes, this approach innovatively introduces an adaptive learnable graph representation matrix to automatically perceive real-time changes in system topologies with no reliance on the knowledge of system topological update information. The main contributions of this paper are as follows.

1) This work develops an ASTGL-based SVS assessment approach that adapts well to various time-varying topological conditions in practical power grids. Compared with existing alternatives relying on exact topological information for SVS feature learning, this approach achieves superior structural change-aware SVS assessment performance via automatic spatiotemporal graph feature learning, being more applicable in practical grids with frequently changing topologies.

2) By integrating adaptive graph representation learning with a spatial attention mechanism, an attention-based graph representation matrix is automatically learned to timely trace the changing spatial features within a specific grid under time-varying topological conditions. Unlike existing GNN-based methods using static or pre-defined graph structures, this dynamic weighting mechanism enables more robust graph representation learning, thereby contributing to a stronger capability in structure-adaptive online SVS assessment.

3) The ASTGL-based SVS assessment model is built by systematically combining spatiotemporal graph convolutional learning with residual learning. While the former manages to mine system-wide spatiotemporal SVS features from diverse structure-varying scenarios, the latter enables stationary in-depth feature learning with smooth gradient propagation. Such a combination results in reliable and stationary SVS assessment against various complicated transient scenarios.

The rest of this paper is organized as follows. Section II comprehensively reviews related works in the literature. The basic problem of GNN-based and TL-based SVS assessment under time-varying topological conditions is described in Section III. Section IV details the proposed ASTGL approach. Section V presents numerical tests on two representative sub-systems of a realistic provincial power grid in South China for verification. Conclusions are finally drawn in Section VI.
\section{Relate Work}
Research on data-driven SVS assessment involves several major directions, ranging from traditional ML-based schemes to deep learning (DL)-enabled solutions, GNN-based methods, and TL-assisted approaches designed to enhance adaptability under topological changes. To provide a systematic review, Table I summarizes the key characteristics and limitations of representative methods in these directions.
\subsection{ML/DL-Based SVS Assessment}
Early data-driven research on SVS assessment primarily can be grouped into two categories. The first one is conventional ML algorithms with shallow learning structures, including DT [4]-[5], RF [6], support vector machine (SVM) [7], extreme learning machine (ELM) [8] and hybrid randomized learning [9], etc. These approaches usually necessitate domain expertise-enabled feature engineering to select proper learning inputs. With the feature engineering procedure heavily relying on domain experts’ subject experiences, the resulting SVS assessment schemes may not always generalize well to practical complex contexts. The other category is DL algorithms with deep neural networks [10], [11], [12]. With no need for customized feature engineering, these solutions can achieve stronger generalization performance via automatic and in-depth SVS feature learning \cite{ref28}. For example, to capture the temporal correlations within post-fault measurement data, a time-adaptive SVS assessment scheme is constructed by adopting the transformer algorithm in [10]. In [11], the transient responsive data collected from various buses in power grids are spatially concatenated in a pixel-like manner, and a CNN is utilized to achieve intelligent SVS assessment. In [12], a spatial attention factor based on static grid topological information is introduced to characterize the spatial correlations between different buses; then, post-fault sequential measurements of individual buses are corrected via the attention factor and fed into a LSTM-based classifier to realize SVS assessment. These DL algorithms have exhibited high SVS assessment performance in respective case studies. However, when applied to practical power grids with irregular network structures, they may fail to capture the grid's inherent non-Euclidean spatial correlations, because their simplified learning mechanisms are generally suited to handling networked data in regular Euclidean space. Consequently, this could deteriorate their performance, being difficult to adapt to complex network structures with irregular spatial correlations.

\begin{table*}[!t]
	\renewcommand{\arraystretch}{1.0}
	\caption{Summary of Representative Stability Assessment Methods}
	\label{tab1}
	\centering
	\footnotesize
	\begin{tabular}{
			>{\centering\arraybackslash}m{0.08\textwidth} % Method Category
			>{\centering\arraybackslash}m{0.06\textwidth} % Reference
			>{\centering\arraybackslash}m{0.10\textwidth} % Reliance on Topological Information
			>{\centering\arraybackslash}m{0.10\textwidth} % Adaptation to Topological Changes
			>{\arraybackslash}m{0.28\textwidth} % Primary Focus
			>{\arraybackslash}m{0.24\textwidth} % Main Limitations
		}
		\hline
		\textbf{Method Category} & \textbf{Reference} & \textbf{Reliance on Topological Information} & \textbf{Adaptation to Topological Changes} & \hspace{45pt}\textbf{Primary Focus} & \hspace{35pt}\textbf{Main Limitations} \\
		\hline
		
		Traditional ML & [4]-[9] & $\times$ & $\times$ &
		$\bullet$ Assess SVS status using shallow ML structures with expert-designed features. &
		$\bullet$ Require heavy domain expertise for feature engineering and limit generalization capability across complex power grid structures. \\
		\hline
		
		DL & [10]-[12] & $\times$ & $\times$ &
		$\bullet$ Learn temporal or spatial SVS features using deep neural architectures. &
		$\bullet$ Fail to handle non-Euclidean spatial correlations effectively. \\
		\hline
		
		GNN & [13]-[20] & $\surd$ & $\times$ &
		$\bullet$ Capture non-Euclidean spatio-temporal correlations for SVS/TSA assessment. &
		$\bullet$ Assume a fixed grid topology and lack adaptability to time-varying network structures. \\
		\hline
		
		TL & [24]-[27] & $\surd$ & $\surd$ &
		$\bullet$ Perform structure-adaptive SVS assessment and mitigating data distribution shift due to topology changes. &
		$\bullet$ Require knowledge of accurate topological change information. \newline $\bullet$ Increase computational burden and limit online applicability due to frequent transfer learning. \\
		\hline
		
		Proposed ASTGL & - & $\times$ & $\surd$ &
		$\bullet$ Perform adaptive spatio-temporal graph learning for SVS assessment. \newline $\bullet$ Real-time perception of spatial features under time-varying topologies. &
		$\bullet$ Increase model complexity due to adaptive graph learning and attention mechanisms. \\
		\hline
	\end{tabular}
\end{table*}
\subsection{GNN-based Dynamic Stability Assessment}
To tackle the limitations of traditional ML/DL in handling non-Euclidean structural data, GNN-based SVS assessment schemes have been introduced in recent studies [13]-[16]. In [13], with post-fault trajectories of individual buses preprocessed via time series shapelet transformation, a graph convolutional network (GCN)-based interpretable SVS assessment model is designed to extract spatiotemporal features from system-wide transient responsive data. Considering that the conventional GCN does not distinguish between different types of buses such as generator buses, tie buses, and load buses during aggregative feature learning, a heterogeneous GNN is introduced in [14] to divide the network graph into a power supply graph and a transmission graph for diversified spatial SVS feature learning. In [15], a spatiotemporal graph convolutional (STGCN)-based SVS classifier is constructed to simultaneously capture the spatial and temporal correlations of adjacent buses. Considering that the assessment performance of a single STGCN model may fluctuate significantly when facing different data acquisition conditions, an ensemble learning strategy is devised in [16] to systematically combine multiple STGCNs for robust SVS assessment.

Concurrently, in the domain of data-driven TSA, GNN models have exhibited high potential as well. In [17], a spatiotemporal broad learning system (STBLS) is constructed by integrating GCN-based spatial modeling and temporal convolutional network (TCN)-based temporal modeling into a broad learning framework, enabling significantly faster TSA training while maintaining competitive accuracy. To enhance a well-trained TSA model’s reliability, a physically interpretable GNN is constructed in [18] to integrate nodal static features, k-hop neighborhood information, and global dependencies captured through a customized self-attention mechanism, thereby alleviating the locality limitation of conventional GNNs. Furthermore, aiming to effectively capture the complex spatiotemporal dynamics inherent in power systems, a GCN-LSTM model is designed in [19] to assess transient power angle stability by capturing topology-aware spatial dependencies and post-disturbance temporal dynamics.

Despite significant progress made in the aforementioned research efforts, they make an implicit assumption that the grid structural information for spatial correlation representation can be preset and remains unchanged across different transient scenarios. Consequently, these may fail to adapt to varying topological structural conditions of practical systems along with time. To address the rigid reliance on a fixed topology, research in the TSA domain has explored graph embedding learning. In [20], a graph embedding-based supervised contrastive learning framework is developed to enhance TSA adaptability under varying grid topologies, where low-dimensional structural embeddings are learned to capture topological differences and improve cross-topology stability prediction. However, a limitation of this method is its focus on capturing only the static topological changes for feature generalization, without leveraging the transient spatiotemporal correlations crucial for SVS assessment.
\subsection{TL-Assisted SVS Assessment}
To realize structure-adaptive SVS assessment, a handful of research efforts have been made via TL schemes [24]-[27]. To reduce the difference between data distributions before and after topological changes, a domain adaptation-based SVS assessment approach is proposed in [24] by minimizing the maximum mean discrepancy. Similarly, in [25], to enhance the adaptability of a trained SVS assessment model to topological changes, transfer learning is carried out to update the model. To enhance the capability of perceiving the power grid structure, a graph attention network-based SVS assessment scheme that can adapt to grid topological changes is introduced in [26] by modeling the adjacency matrix in topological updated scenarios. Considering the sparsity of power grid topologies, a topology adaptive reconstruction-based transfer learning scheme for SVS assessment is proposed in [27]. These studies have exhibited high potential to enhance the adaptability of data-driven SVS assessment schemes to topological variations. However, when faced with time-varying topological conditions in practice, transfer learning necessitates the generation of a small number of new samples under the changed topological structure for learning model updates. This implies that the reliability of these schemes largely depends on the knowledge of the accurate topological change information during online monitoring. Besides, if system topologies change frequently in practice, transfer learning needs to be implemented repeatedly, thereby increasing the computational burden during online application. Consequently, their reliability and applicability may be restricted in practical contexts.
\section{Problem Description}
Given a specific power system containing \emph{N} buses for SVS monitoring, a GNN-based SVS assessment model {\bf{\emph{G}}} can be constructed by learning spatio-temporal correlations between individual buses across various transient operation scenarios:
\begin{equation}
	\fontsize{9pt}{9pt}\selectfont
	\boldsymbol{G}: (\boldsymbol{X}, \boldsymbol{A}) \rightarrow y, \text{ for } \boldsymbol{X} = \{\boldsymbol{x}_1, \boldsymbol{x}_2, \ldots, \boldsymbol{x}_N\}, \ y = \{0, 1\}
\end{equation}
where \(\boldsymbol{x}_i \!\in\! \mathbb{R}^{L \times F}\) denotes \(F\)-dimensional time series (with \(L\) data points in total) acquired from monitored bus \(i\) after a specific transient fault. \(\boldsymbol{A} \!\!\in\!\! \mathbb{R}^{N \times N}\) denotes the connectivity relationships between monitored buses, i.e., the adjacency matrix. The element \(a_{ij}\) in the matrix \(\boldsymbol{A}\) is specified as \(a_{ij} \!=\! 1\) or \(1/z_{ij}\) if there is a connection between monitored buses \(i\) and \(j\), and \(a_{ij} \!=\! 0\) otherwise [13]-[20]. Here \(z_{ij}\) represents the impedance of the transmission line connecting buses \(i\) and \(j\). \(y\) indicates the system SVS status at the end of the transient process (\(0 \!\rightarrow\! \text{stable}\), \(1 \!\rightarrow\! \text{unstable}\)). In general, taking data \(\boldsymbol{X}\), \(\boldsymbol{A}\), and \(y\) as input-output pairs for advanced GNN-based representation learning can help the system achieve reliable SVS assessment from a structural data-driven perspective.

However, as shown in Fig. 1, in the presence of continuous topological changes during online monitoring, it is difficult for a conventional GNN model to trace and adapt to such changes along with time, thereby inevitably degrading the SVS assessment performance. Essentially, the degraded SVS assessment primarily results from the GNN model’s heavy reliance on an initially preset adjacency matrix \(\boldsymbol{A}\), which makes the learned model less effective in adapting to the changes of the system topology during online application. Furthermore, while TL-based schemes have been explored to achieve structure-adaptive SVS assessment, they suffer from significant limitations when facing time-varying topologies in practice. Specifically, TL requires the generation of new samples and the repeated implementation of model updates for each new topological scenario, imposing a high computational burden and requiring prior knowledge of the exact topological change information during online monitoring. Consequently, the reliability and applicability of both fixed-topology GNNs and new samples-dependent TL schemes are restricted in practical contexts characterized by frequent and dynamic topological variations.
\begin{figure}[!t]
	\centering
	\includegraphics[width=3.5in]{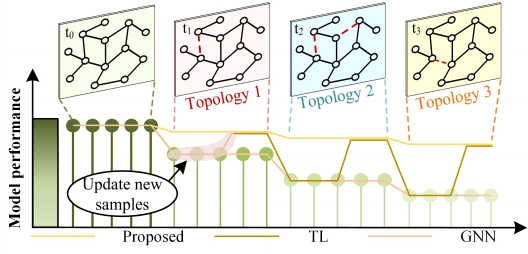}
	\caption{Illustration of topological changes in a specific power system.}
	\label{fig1}
\end{figure}
\begin{figure}[!t]
	\centering
	\includegraphics[width=3.5in]{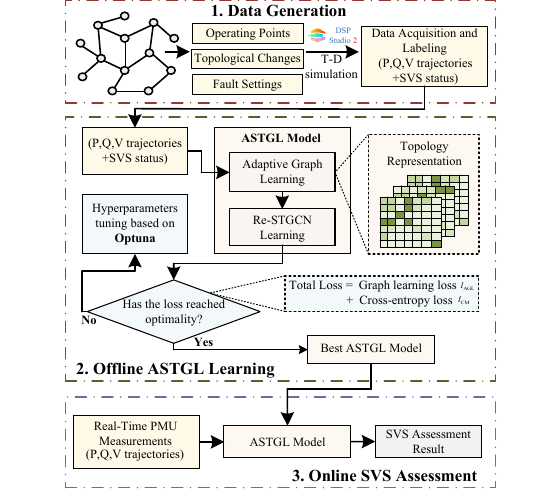}
	\caption{Overall framework of the proposed ASTGL approach.}
	\label{fig2}
\end{figure}

To address the aforementioned issue, this paper proposes an ASTGL approach that can adaptively explore the system’s topological representation under time-varying topological conditions and more effectively extract SVS features from the irregular networked spatio-temporal perspective. Specifically, the primary spatial correlations between individual buses within the system of interest are inferred based on the similarities of their transient responses. Then, the inferred correlations are mapped to weights in the graph representation matrix through a learnable parametric function. Along with subsequent SVS assessment model training, these weights would be automatically adjusted according to the transient responses of different SVS cases, thereby ensuring that the matrix can flexibly adapt to the changes of the system topology. In order to help robustly learn system SVS features from the spatial perspective, a spatial attention mechanism is further introduced to dynamically adjust the graph representation matrix under different transient scenarios. Based on the adjusted matrix, a residual-STGCN (Re-STGCN) is introduced to comprehensively perform spatiotemporal SVS feature learning. By integrating residual connections with spatiotemporal graph convolutional operations, the learning model realizes structure-adaptive SVS assessment with salient robustness and generalization capability. Furthermore, the Optuna-based optimization \cite{ref29} is introduced into the learning procedure to help automatically tune hyperparameters for the ASTGL. More details about the whole approach are presented in Section IV.
\section{Proposed Methodology}
Overall framework of the proposed ASTGL for online SVS assessment is shown in Fig. 2. Its implementation includes three phases: 1) data generation; 2) offline ASTGL learning; and 3) online SVS assessment.
\subsection{Data Generation}
First, various generation and loading conditions as well as possible topological changes are considered to construct a representative operation set that widely covers diverse operation points with typical system topological variations. For each operating point, batch time-domain (TD) simulations are carried out to mimic various transient scenarios, involving different fault types, fault locations and fault durations. All of these simulations result in an initial SVS case base with \(s\) cases. For each case, voltage magnitude acting as the direct indicator of system SVS levels is taken as the major input features for subsequent SVS assessment model derivation. In addition, the variations of the active and reactive power that inherently induce voltage fluctuations during SVS evolution are taken as input features as well. Therefore, the time-series data of active power, reactive power, and voltage magnitude, i.e., $\{P, Q, V\}$, of individual monitored buses are taken as the input of the ASTGL model, where $P$ and $Q$ denote the active and reactive power injections at each bus. Specifically, assuming that the fault is cleared at the time instant $t_c$, an observation time window (OTW) with time length \(T_{\text{win}}\) and sampling interval $\Delta t$ ($T_{\text{win}} \!=\! L \!\times\! \Delta t$) is set to acquire transient responsive trajectories $\{P, Q, V\}$ from each monitored bus within the time span of [$t_c$, $t_c$+\(T_{\text{win}}\)].

For system stability status $y$, this paper adopts the criterion widely applied in China for SVS determination \cite{ref30}: after a transient fault, if all the load-bus voltages in a specific system recover to more than 0.8 pu within 10 s, the system SVS status is deemed as stable ($y = 0$), otherwise unstable ($y = 1$). By collecting the initial transient trajectories \(\boldsymbol{X} \!\in\! \mathbb{R}^{L \times N \times 3}\) and eventual SVS status $y$ of all the SVS cases, an SVS sample set $S = \{(\boldsymbol{X}_i, y_i) \mid \boldsymbol{X}_i = (x_1, x_2, \ldots, x_N), 1 \leq i \leq s\}$ is constructed for subsequent ASTGL model training.
\subsection{Offline ASTGL Learning}
As shown in Fig. 3, the ASTGL model structure consists of four key modules: 1) adaptive graph learning module (AGLM), 2) spatial attention module (SAM), 3) Re-STGCN module and 4) classification module (CM). Each module will be illustrated in detail as follows.
\begin{figure}[!t]
	\centering
	\includegraphics[width=3.5in]{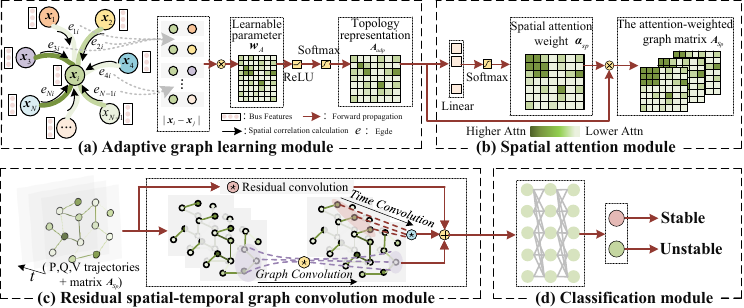}
	\caption{Proposed ASTGL network structure.}
	\label{fig3}
\end{figure}

\textit{1) Adaptive Graph Learning Module}: To adapt to the variable topological conditions, this paper designs an AGLM, inspired by the work in \cite{ref31}. Due to the inherent network electrical coupling between individual buses, buses with close spatial distances generally exhibit highly similar evolution trends after transient faults. Hence, the transient responsive trajectories of the monitored buses in the system, i.e., $\boldsymbol{X}$, can be used to adaptively capture the spatial correlations between individual buses under different topological conditions. Specifically, we define a non-negative learnable element $e_{ij}$ to represent spatial correlation between bus $i$ and bus $j$ in the power grid. With no requirement for the knowledge of exact topological information, $e_{ij}$ can be obtained through iterative training of the learnable parameter matrix $\boldsymbol{w}_{ij}^A \in \mathbb{R}^{F \times L}$:
\begin{equation}
	\fontsize{10pt}{10pt}\selectfont
	e_{ij} = \frac{1}{L} \times \frac{\sum_{t=1}^{L} \exp(-\text{ReLU}(\boldsymbol{w}_{ij}^A \mid \boldsymbol{x}_i - \boldsymbol{x}_j \mid))}{\sum_{t=1}^{L} \sum_{j=1}^{N} \exp(-\text{ReLU}(\boldsymbol{w}_{ij}^A \mid \boldsymbol{x}_i - \boldsymbol{x}_j \mid))}
\end{equation}
\begin{equation}
	\fontsize{10pt}{10pt}\selectfont
	\boldsymbol{A}_{\text{adp}} = \begin{bmatrix}
		e_{11} & e_{12} & \cdots & e_{1N} \\
		e_{21} & e_{22} & \cdots & e_{2N} \\
		\vdots & \vdots & \ddots & \vdots \\
		e_{N1} & e_{N2} & \cdots & e_{NN}
	\end{bmatrix}
\end{equation}
where ReLU denotes the nonlinear activation function, which ensures the nonnegativity of $e_{ij}$. $\boldsymbol{x}_i \!=\! \{P_i, Q_i, V_i\} \in \mathbb{R}^{L \times 3}$ represents the transient responsive trajectories of bus $i$. $L$ is the number of data points. $\exp(\cdot)$ is the exponential function. $\boldsymbol{A}_{\text{adp}} \!\in\! \mathbb{R}^{N \times N}$ is the learned graph matrix characterizing the topology of the power grid. Based on the $\exp(\cdot)$ operation, $e_{ij}$ falls into the range of [0,1]. The larger the value of $e_{ij}$, the stronger the spatial correlation between buses $i$ and $j$.

To ensure the effectiveness of AGLM under variable topological conditions, a graph learning loss function is designed to dynamically adjust each element in the adaptive matrix:
\begin{equation}
	\fontsize{10pt}{10pt}\selectfont
	\boldsymbol{l}_{\text{AGL}} = \sum_{i,j=1}^{N} \left\| \boldsymbol{x}_i - \boldsymbol{x}_j \right\|_2^2 e_{ij} + \lambda \left\| \boldsymbol{A}_{\text{adp}} \right\|_F^2
\end{equation}
where $\left\| \boldsymbol{x}_i - \boldsymbol{x}_j \right\|_2^2$ represents the square of the Euclidean distance between bus $i$ and bus $j$ in the embedding space. By minimizing $\left\| \boldsymbol{x}_i - \boldsymbol{x}_j \right\|_2^2 e_{ij}$, this ensures that buses with similar responsive trajectories are mapped to the closest embedding space locations during the graph learning process. $\left\| \boldsymbol{A}_{\text{adp}} \right\|_F$ denotes the Frobenius norm of the learnable representation matrix. $\lambda \geq 0$ is the regularization coefficient. Minimizing $\lambda \left\| \boldsymbol{A}_{\text{adp}} \right\|_F^2$ prevents the elements in $\boldsymbol{A}_{\text{adp}}$ from becoming excessively large, thereby maintaining numerical stability. By doing so, this can reduce the model’s computational complexity, enhance the interpretability of the adaptive matrix, and ensure that the AGLM can generalize well to variable topological conditions.

\textit{2) Spatial Attention Module}: The attention mechanism, by simulating the human brain's attention, adaptively assigns different weights to different input features and reduces information redundancy \cite{ref32}. During the post-fault transient responsive process, reactive power cannot be transmitted over long distances and the voltage evolution of each bus primarily affects its electrically adjacent buses. In this context, voltage instability regions with severe voltage depression often present spatially local distributions. For the adaptive matrix $\boldsymbol{A}_{\text{adp}}$, although it preliminarily realizes grid structural representation under time-varying topological conditions, it is difficult to capture these local distributions. To address this inadequacy, the SAM is further introduced to dynamically adjust the weights in $\boldsymbol{A}_{\text{adp}}$. The network structure of SAM is shown in Fig. 3(b). Specifically, taking $\boldsymbol{A}_{\text{adp}}$ as an example, the spatial attention weight matrix $\boldsymbol{Z}_{\text{sp}} \!=\! (\boldsymbol{z}_1, \boldsymbol{z}_2, \ldots, \boldsymbol{z}_N) \!\in\! \mathbb{R}^{N \times N}$ can be calculated by a single-layer neural network:
\begin{equation}
	\fontsize{10pt}{10pt}\selectfont
	\boldsymbol{Z}_{\text{sp}} = \sigma(\boldsymbol{w}^{\text{sp}} \boldsymbol{A}_{\text{adp}} + \boldsymbol{b}^{\text{sp}})
\end{equation}
where \(\boldsymbol{w}^{\text{sp}} \in \mathbb{R}^{N \times N}\) is the learnable parameter matrix. \(\boldsymbol{b}^{\text{sp}} \!\in\! \mathbb{R}^{N}\) is the bias vector. \(\sigma(\cdot)\) is the sigmoid activation function. Then, by normalizing the weight matrix \(\boldsymbol{Z}_{\text{sp}}\), the spatial attention score $\boldsymbol{\alpha}_{\text{sp}} \!=\! (\boldsymbol{\alpha}_1^{\text{sp}}, \boldsymbol{\alpha}_2^{\text{sp}}, \ldots, \boldsymbol{\alpha}_N^{\text{sp}}) \!\in\! \mathbb{R}^{N \times N}$ between individual buses in \(\boldsymbol{A}_{\text{adp}}\) can be obtained by
\begin{equation}
	\fontsize{10pt}{10pt}\selectfont
	\boldsymbol{\alpha}_i^{sp} = \frac{\exp(\boldsymbol{z}_i)}{\sum_{l=1}^{N} \boldsymbol{z}_{i,l}}
\end{equation}
where $\boldsymbol{\alpha}_i^{\text{sp}} \!\in\! \mathbb{R}^{N}$ is the attention score between bus $i$ and other buses, with the range $[0,1]$, and $\sum_{l=1}^{N} \boldsymbol{\alpha}_{i,l} = 1$. By weighting and summing each element in $\boldsymbol{A}_{\text{adp}}$ according to the attention score $\boldsymbol{\alpha}^{\text{sp}}$, the grid representation matrix $\boldsymbol{A}_{\text{sp}}$ after SAM's attention is obtained by
\begin{equation}
	\fontsize{10pt}{10pt}\selectfont
	\boldsymbol{A}_{\text{sp}} = \boldsymbol{\alpha}^{\text{sp}} \odot \boldsymbol{A}_{\text{adp}} = [\boldsymbol{\alpha}_1^{\text{sp}} \boldsymbol{e}_1, \boldsymbol{\alpha}_2^{\text{sp}} \boldsymbol{e}_2, \ldots, \boldsymbol{\alpha}_N^{\text{sp}} \boldsymbol{e}_N]
\end{equation}
Here the attention score $\boldsymbol{\alpha}^{\text{sp}}$ is utilized to assign greater weights to voltage instability regions with severer voltage sags. The large elements within $\boldsymbol{\alpha}^{\text{sp}}$ can automatically guide the whole SVS assessment approach to pay more attention to regions with severe SVS problems. By doing so, the attention-based grid representation matrix $\boldsymbol{A}_{\text{sp}}$ can contribute to more robust SVS feature learning from a regional perspective, thereby helping the whole SVS assessment approach better adapt to changing topological conditions.

\textit{3) Re-STGCN Learning Module}: To enhance spatiotemporal correlation learning from different SVS cases, a residual convolutional connection is introduced in parallel with a STGCN. The network structure of Re-STGCN is shown in Fig. 3(c). Firstly, with the grid characterized by the grid representation matrix $\boldsymbol{A}_{\text{sp}}$, the graph Laplacian matrix $\boldsymbol{J}$ of the system is normalized as follows:
\begin{equation}
	\fontsize{10pt}{10pt}\selectfont
	\boldsymbol{J} = \boldsymbol{I} - \boldsymbol{D}^{-1/2} \boldsymbol{A}_{\text{sp}} \boldsymbol{D}^{-1/2}
\end{equation}
where \(\boldsymbol{I} \!\in\! \mathbb{R}^{N \times N}\) is the identity matrix, \(\boldsymbol{D} \!\in\! \mathbb{R}^{N \times N}\) is a diagonal matrix with \(D_{ii} = \sum_j e_{ij}^{\text{sp}}\). Then, to aggregate the spatial features of buses with close electrical distances, with the transient response \(\boldsymbol{X}_t = [P_t, Q_t, V_t] \in \mathbb{R}^{N \times 3}\), the spatial graph convolution of STGCN can be manipulated by approximating Chebyshev polynomials \cite{ref33}:
\begin{equation}
	\fontsize{10pt}{10pt}\selectfont
	\boldsymbol{Z}_j(t) = 
	\begin{cases} 
		\Theta *_G \boldsymbol{P}_t \approx \sum_{j=1}^{3} \sum_{k=0}^{K_s-1} \theta_{k,j} T_k(\tilde{\boldsymbol{J}}) \boldsymbol{P}_t \\
		\Theta *_G \boldsymbol{Q}_t \approx \sum_{j=1}^{3} \sum_{k=0}^{K_s-1} \theta_{k,j} T_k(\tilde{\boldsymbol{J}}) \boldsymbol{Q}_t \\
		\Theta *_G \boldsymbol{V}_t \approx \sum_{j=1}^{3} \sum_{k=0}^{K_s-1} \theta_{k,j} T_k(\tilde{\boldsymbol{J}}) \boldsymbol{V}_t
	\end{cases}
\end{equation}
where \(\boldsymbol{Z}_j(t)\) is the graph convolution result for the \(j\)-th feature channel (\(1 \leq j \leq 3\)). \(\Theta \in \mathbb{R}^{K_s \times 3 \times f_{\text{gcn}}}\) represents the spatial filtering matrix of GCN. \(K_s\) is the Chebyshev polynomial order. \(f_{\text{gcn}}\) is the number of output channels. \(*_G\) denotes the spectral GCN operation. \(\theta_{k,j}\) is the learnable parameter of the \(k\)-th order (\(0 \leq k \leq K_s\)). \(\tilde{\boldsymbol{J}} = 2\boldsymbol{J}/\lambda_{\max} - \boldsymbol{I}\) is the scaled eigenvector matrix. \( \lambda_{\max} \) is the largest eigenvalue of the Laplacian matrix \( \boldsymbol{J} \). Here \( T_k(\tilde{\boldsymbol{J}}) \) is the Chebyshev polynomial of the \( k \)-th order:
\begin{equation}
	\fontsize{10pt}{10pt}\selectfont
	T_k(\tilde{\boldsymbol{J}}) = 
	\begin{cases} 
		1, & k=0 \\
		\tilde{\boldsymbol{J}}, & k=1 \\
		2\tilde{\boldsymbol{J}} T_{k-1}(\tilde{\boldsymbol{J}}) - T_{k-2}(\tilde{\boldsymbol{J}}), & 2 \leq k \leq K_s
	\end{cases}
\end{equation}

Further, the transient responsive trajectories \( \boldsymbol{X} \in \mathbb{R}^{L \times N \times 3} \) are input into Eq. (9). Since \( K_s \) represents the receptive field of graph convolution, the spatial feature information of multi-order neighbor buses over the OTW can be aggregated as \( \boldsymbol{Z}_s = \Theta * \boldsymbol{X} \), with \( \boldsymbol{Z}_s \in \mathbb{R}^{L \times N \times f_{\text{gcn}}} \). Then, on the basis of \(\boldsymbol{Z}_s\), the time convolution (TCN) manipulation of STGCN can be formulated as follows:
\begin{equation}
	\fontsize{10pt}{10pt}\selectfont
	\boldsymbol{Z}_{\text{st}} = \text{ReLU}\left(\Phi * \left(\text{ReLU}\left(\boldsymbol{Z}_\text{s}\right)\right)\right)
\end{equation}
where \(\boldsymbol{Z}_{\text{st}} \in \mathbb{R}^{f_{\text{tcn}} \times N \times f_{\text{gcn}}}\) denotes the spatio-temporal features extracted from \( \boldsymbol{X} \). \( \Phi\!\!\in\!\!\mathbb{R}^{K_t \times L \times f_{\text{tcn}}} \) is the TCN's filter matrix. \(K_t\) is the convolution kernel size. \( f_{\text{tcn}} \) is the number of output channels for TCN filtering. * denotes the temporal convolution operation. As the depth of stacked STGCN layers increases, the model's gradients may gradually vanish during backpropagation, leading to slow convergence of spatiotemporal learning. To address this issue, residual convolution in Re-STGCN is introduced to allow gradient propagation from the shallow layers to the deep layers of the network:
\begin{equation}
	\fontsize{10pt}{10pt}\selectfont
	\boldsymbol{Z}_\text{st}^\prime = \text{ReLU}\left(\Phi * \left(\text{ReLU}\left(\boldsymbol{X}\right)\right) + \boldsymbol{Z}_\text{st} \right) \tag{12}
\end{equation}
where \(\boldsymbol{Z}_\text{st}^\prime \in \mathbb{R}^{f_{\text{tcn}} \times N \times f_{\text{gcn}}}\) denotes the enhanced spatiotemporal features via residual convolution. By combining the initial responsive trajectories and the STGCN layer's output, the residual connections improve the whole model's spatiotemporal learning convergence, thereby avoiding model degradation caused by the stacking of STGCN layers.

\textit{4) Classification Module}: As the last module of the ASTGL model, the CM uses the fully-connected layer to reduce the dimensionality of the spatiotemporal features extracted by Re-STGCN module, and then maps the reduced features to a probability distribution via the softmax function:
\begin{equation}
	\fontsize{10pt}{10pt}\selectfont
	\hat{\boldsymbol{y}} = \text{softmax}(\boldsymbol{w}^{cm}\boldsymbol{Z}_\text{st}^\prime + \boldsymbol{b}^{\text{cm}}) \tag{13}
\end{equation}
where \(\boldsymbol{w}^{\text{cm}}\) denotes the learnable parameter. \(\boldsymbol{b}^{\text{cm}} \in \mathbb{R}^2\) is the bias vector. \(\hat{\boldsymbol{y}} \in (0,1)\) is the SVS prediction probability, and \(\sum_{i=1}^{2} \hat{\boldsymbol{y}}_i = 1\). Given that SVS assessment is a binary classification task, this paper selects the cross-entropy loss function as the classification loss for the ASTGL model:
\begin{equation}
	\fontsize{10pt}{10pt}\selectfont
	\boldsymbol{l}_{\text{CM}} = -y_i \log \hat{y}_i - (1 - y_i) \log (1 - \hat{y}_i) \tag{14}
\end{equation}
where \( y_i \) is the true SVS status for the \( i \)-th case. Combining the graph learning loss in Eq. (4), the total learning loss of the ASTGL model is as follows:
\begin{equation}
	\fontsize{10pt}{10pt}\selectfont
	\boldsymbol{l}_{\text{Total}} = \boldsymbol{l}_{\text{AGL}} + \boldsymbol{l}_{\text{CM}} \tag{15}
\end{equation}
Here, \(\boldsymbol{l}_{\text{AGL}}\) dynamically adjusts edge weights in the graph representation matrix to preserve the topological properties of the grid structure. Concurrently, \(\boldsymbol{l}_{\text{CM}}\) prompts Re-STGCN to efficiently extract spatiotemporal features strongly correlated with SVS. Through backpropagation \cite{ref34}, model parameters are gradually updated to minimize the total loss \(\boldsymbol{l}_{\text{Total}}\), a process that ultimately guides the ASTGL model to realize structure-adaptive SVS assessment. Note that, to fully unlock the potential of the ASTGL model, an Optuna-based optimization \cite{ref29} is introduced into the learning procedure to help automatically tune hyperparameters for the ASTGL model. As an automated hyperparameter optimization framework, Optuna can iteratively call and evaluate the ASTGL model with different hyperparameter sets to obtain the optimal solution. In the sequel, along with the description of specific case studies, the detailed Optuna-based optimization process will be introduced in Section V-B.
\subsection{Online SVS Assessment}
During online SVS monitoring under time-varying topological conditions, when the system encounters a transient event, a measured data matrix $\boldsymbol{X}_{\text{mea}}$ with the OTW \(T_{\text{win}}\) and sampling interval $\Delta t$ is acquired from different monitored buses by phasor measurement units (PMUs). Subsequently, $\boldsymbol{X}_{\text{mea}}$ is fed into the well-trained ASTGL model to predict the system's SVS status. Whether to take control measures is determined by the SVS assessment result. If the prediction indicates stability, the OTW slides to the next time period for continuous SVS monitoring. Otherwise, alarm signals would be triggered immediately to warn that corresponding emergency measures should be taken as soon as possible.

To comprehensively test the SVS assessment performance of the ASTGL model, four evaluation metrics are selected, including: Accuracy (\textit{Acc}), F1 score (\textit{F1}), Recall (\textit{Rec}), and Precision (\textit{Pre}). Their definitions are as follows:
\begin{equation}
	\fontsize{10pt}{10pt}\selectfont
	Acc = \frac{TP + TN}{TP + TN + FN + FP} \times 100\% \tag{16}
\end{equation}
\begin{equation}
	\fontsize{10pt}{10pt}\selectfont
	F1 = \frac{2 \cdot Pre \cdot Rec}{Pre + Rec} \tag{17}
\end{equation}
\begin{equation}
	\fontsize{10pt}{10pt}\selectfont
	Pre = \frac{TP}{TP + FP} \tag{18}
\end{equation}
\begin{equation}
	\fontsize{10pt}{10pt}\selectfont
	Rec = \frac{TP}{TP + FN} \tag{19}
\end{equation}
where \textit{TP} represents the number of correctly recognized unstable samples, \textit{TN} represents the number of correctly classified stable samples, \textit{FP} represents the number of falsely alarmed stable samples, and \textit{FN} represents the number of misclassified unstable samples.
\section{Case Study}
To demonstrate the performance of the proposed ASTGL approach, a series of comprehensive tests were conducted on a power grid in a southern province of China with 66 buses at the 500-kV level and 1368 buses at 220-kV and lower voltage levels. As illustrated in Fig. 4 (for clarity, only the 500-kV and 220-kV backbone networks are shown), the system represents the initial topological configuration of the study area. It consists of regular power plants and multiple DC transmission lines, with a total installed capacity of 74 GW. All power generators are equipped with excitation systems to enhance voltage stability.
 
As the main concern of this paper is the SVS problem against time-varying network structures, two sub-systems in the southeast region of this power grid are selected for regional SVS monitoring. Specifically, simulation configuration and case generation under topological variation scenarios are first conducted on sub-system 1, highlighted in light green in Fig. 4, which consists of 69 buses at 500-kV and lower voltage levels. Subsequently, automated hyperparameter optimization of the proposed ASTGL model is performed using Optuna, followed by an analysis of the influence of key parameters on SVS evaluation performance. The test results obtained from Sub-system 1 are then compared with those of several state-of-the-art methods. Finally, to validate the scalability of the proposed approach in large-scale systems, sub-system 2, highlighted in dark green in Fig. 4 and comprising 319 buses at 500-kV and lower voltage levels, is employed for comparative analysis. System component modeling and batch-TD simulations were conducted using a commercial Root Mean Square (RMS) electromechanical simulation package called DSP-Studio, which is developed by the Electric Power Research Institute of China Southern Power Grid. The induction motors were represented by a fifth-order dynamic model, with their mechanical torque characterized by a quadratic torque-speed relationship consistent with typical industrial load characteristics.
\subsection{Simulation Setting and Case Generation}
The sub-system 1’s 500-kV and 220-kV backbone structure covering five major regions is illustrated in Fig. 4. For the sake of sufficiently mimicking diverse topological changes, different 220-kV transmission lines in the system are randomly tripped to generate three distinct scenarios.
\begin{figure}[!t]
	\centering
	\includegraphics[width=3.5in]{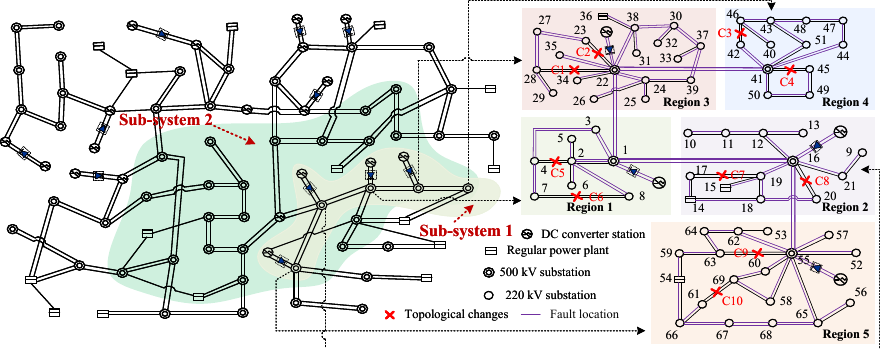}
	\caption{Backbone structure of a realistic province power grid in South China.}
	\label{fig4}
\end{figure}
\begin{figure}[!t]
	\centering
	\includegraphics[width=3.5in]{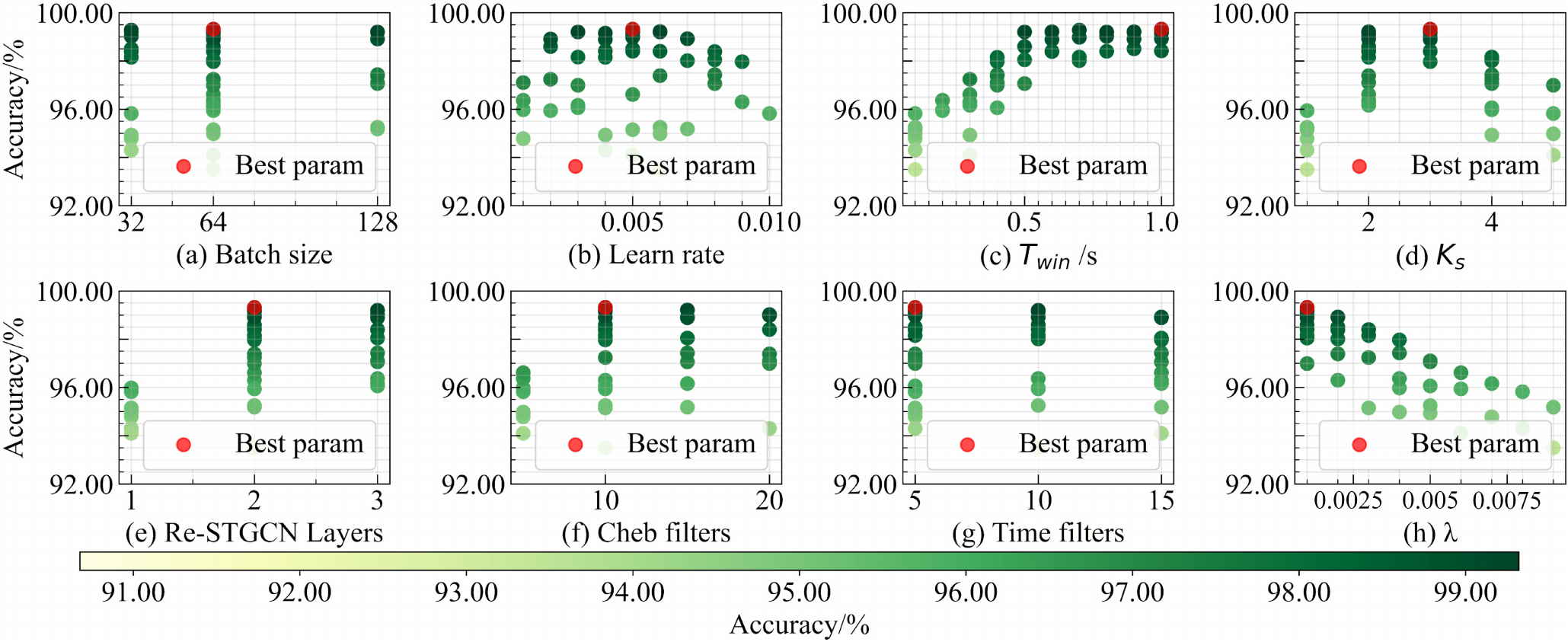}
	\caption{Proposed ASTGL hyperparameter results based on TPE Bayesian optimization (red point denotes the optimal configuration).}
	\label{fig5}
\end{figure}

1) Scenario A: It only contains one topological condition with the base structural configuration, as shown in Fig. 4.

2) Scenario B: With five typical 220-kV transmission lines (C1, C3, C5, C7, C9) within the five regions, we stochastically trip one or two of these lines to generate \(N\)-1 and \(N\)-2 topological variations.

3) Scenario C: Five representative 220-kV transmission lines (C2, C4, C6, C8, C10) are chosen, and one or two of these lines are randomly tripped to generate \(N\)-1 and \(N\)-2 topological variations.

For each topological change scenario, the following settings are considered: 1) The load level is altered within the range of 80\% to 120\% of the system's base load to generate different operating points. Accordingly, the power generation of the individual generating units is adjusted to ensure new power balance. 2) All loads in the target system are configured in the form of induction motor connected in parallel with static ZIP loads. The proportion of induction motor is set to \{0.6, 0.75, 0.9, 0.95\}. 3) Typical AC faults and DC faults are simulated in the five regions, as illustrated by the locations of the purple transmission lines in Fig. 4. The AC faults involve three-phase \(N\)-1 and \(N\)-2 short-circuit faults, with fault duration and fault location set to \{0.1 s, 0.2 s, 0.3 s\}\ and \{2\%, 50\%, 98\%\}, respectively. The DC faults consider types such as bipolar blocking and commutation failure.

Combining the above three groups of topological variations (i.e., Scenarios A, B and C) and simulations settings, three datasets are obtained from batch-TD simulations. The generated datasets are statistically summarized in Table II. In this paper, all the buses in the five regions are taken as the SVS monitoring sites (a total of 69 buses), and transient responsive trajectories of $\{P, Q, V\}$ are acquired from these buses after the transient fault clearance (with the OTW \(T_{\text{win}}\)=0.5 s and sampling interval $\Delta t$=0.01 s). For datasets A and B, their topological conditions are assumed to be known, and they are used for training the proposed ASTGL model and other comparative methods in the sequel like STGCN (see Section V-C.2) which requires the knowledge of the grid topological structure. For dataset C, it is assumed that the topological structures are totally unknown, which is used to verify the generalization performance under unknown topological scenarios. 
\begin{table}[!t]
	\centering
	\renewcommand{\arraystretch}{1.0}
	\caption{Statistics of Generated Datasets}
	\label{tab2}
	% 使用 resizebox 将表格宽度严格限制在当前栏宽内，防止超出版面
	\resizebox{\linewidth}{!}{
		\begin{tabular}{>{\centering\arraybackslash}m{3.2cm} >{\centering\arraybackslash}m{1cm} >{\centering\arraybackslash}m{1.5cm} >{\centering\arraybackslash}m{1.5cm} >{\centering\arraybackslash}m{1.2cm}}
			\toprule
			\multirow{2}{=}{\centering Dataset} & \multirow{2}{=}{\centering Split} & \multicolumn{2}{c}{Labels} & \multirow{2}{=}{\centering Total} \\ \cmidrule{3-4}
			& & Stable & Unstable & \\ \midrule
			
			\multirow{2}{=}{\centering A (One topology)} & Train & 3064 & 3064 & 6128 \\
			& Test & 766 & 766 & 1532 \\ \midrule
			
			\multirow{2}{=}{\centering B (Topological changes)} & Train & 3077 & 3078 & 6155 \\
			& Test & 770 & 769 & 1539 \\ \midrule
			
			C (Topological changes) & Test & 770 & 769 & 1539 \\
			\bottomrule
		\end{tabular}
	}
\end{table}
\subsection{Hyperparameter Setting of the ASTGL Model}
For SVS assessment under topological change scenarios, several key hyperparameters exhibit substantial sensitivity to the performance of the ASTGL model. To a balanced configuration, this paper adopts the Optuna-based Bayesian optimization framework on dataset B to optimize the hyperparameters of the proposed ASTGL model. The optimization results are shown in Fig. 5, where each trial corresponds to a specific hyperparameter configuration and the red marker denotes the optimal one. As observed, Optuna effectively identifies the high-performance regions in the parameter space, achieving a best validation accuracy of 99.37\%. Among all the searched parameters, \(T_{\text{win}}\), $\lambda $ and \(K_s\) exhibited the most prominent influence on accuracy of the SVS assessment. Specifically, \(T_{\text{win}}\) controls the effective temporal receptive field of the Re-STGCN block, while $\lambda $ and \(K_s\) jointly determine the balance between the learned adaptive topology and the local spatial receptive capacity. Hence, their effects are further analyzed in the following subsections.

\textit{1) Effect of \(T_{\text{win}}\)}: A shorter OTW length \(T_{\text{win}}\) enables earlier SVS assessment, allowing the system more time to implement emergency measures. Therefore, this subsection evaluates the sensitivity to the OTW length \(T_{\text{win}}\) by training and testing ASTGL with transient response sequences of varying lengths, while keeping all other hyperparameters fixed at their optimal values. The corresponding results are presented in Fig. 6(a).

As shown in Fig. 6(a), the SVS assessment accuracy of the ASTGL model increases with the OTW length. The accuracy reaches 99.22\% at \(T_{\text{win}} = 0.5s\) and peaks at 99.37\% when \(T_{\text{win}} = 1.0s\). This indicates that extending \(T_{\text{win}}\) allows the model to capture richer temporal dependencies by incorporating more historical information. However, an excessively long \(T_{\text{win}}\) may delay the SVS assessment and hinder timely emergency measures. Considering this trade-off, \(T_{\text{win}} = 0.5s\) is adopted in this study, as it provides sufficient time for emergency response while maintaining high accuracy.

\textit{2) Influence of $\lambda $ and \(K_s\)}: Fig. 6(b) illustrates how the joint variation of $\lambda $ and \(K_s\) affects the SVS assessment accuracy achieved by the proposed ASTGL model. As observed, when $\lambda $ is extremely small, the AGLM retains high flexibility, allowing the model to capture dynamic spatial dependencies effectively. However, increasing $\lambda $ excessively enforces stronger smoothness constraints on the learned topology, which reduces its adaptability and leads to degraded performance across all \(K_s\) values. Meanwhile, for a fixed $\lambda $, the accuracy first increases \(K_s\) and then gradually decreases, indicating that moderate receptive field expansion enhances spatial feature extraction, whereas overly large \(K_s\) introduces redundant or noisy information. Consequently, the optimal configuration, i.e., $\lambda=0.0001$ and \(K_s\)=3 achieves the highest accuracy 99.22\%, balancing adaptive topology learning and local spatial aggregation for time-varying network structures.

According to the above analysis, the optimal structural configuration of the ASTGL model is provided in a downloadable supplementary file \cite{ref35} (see Table S2 in the file).

\begin{figure}[!t]
	\centering
	\includegraphics[width=3.5in]{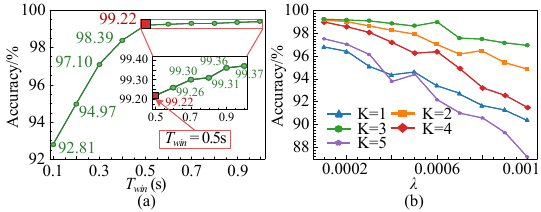}
	\caption{Sensitivity analysis of key hyperparameters on SVS assessment accuracy of the proposed ASTGL model: (a) Effect of the OTW length \(T_{\text{win}}\), (b) Joint influence of $\lambda $ and \(K_s\).}
	\label{fig6}
\end{figure}
\begin{figure}[!t]
	\centering
	\includegraphics[width=3.5in]{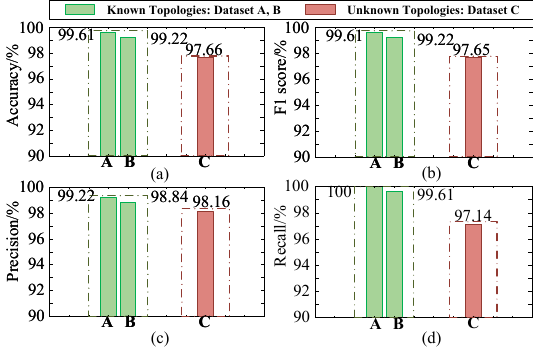}
	\caption{SVS assessment performance of the proposed ASTGL under known/unknown topological changes. (a) \textit{Acc}, (b) \textit{F1}, (c) \textit{Pre}, (d) \textit{Rec}.}
	\label{fig7}
\end{figure}
\subsection{Comprehensive Performance Verification}
\textit{1) Overall Performance on SVS Assessment}: The comprehensive SVS assessment results of the proposed ASTGL approach on the three datasets are shown in Fig. 7. It can be seen that the proposed approach achieves excellent online SVS assessment results under both known and unknown topological changes. For dataset A involving only one topological condition, the overall SVS assessment accuracy \textit{Acc} of the ASTGL model is as high as 99.66\%. Additionally, its \textit{Rec} index reaches 100\%, indicating that the ASTGL model can reliably identify all the unstable cases. Further, for dataset B involving multiple topological change conditions, the online SVS assessment results of the ASTGL model are slightly affected, with \textit{Acc} decreasing by no more than 0.5\%. Notably, by checking the assessment results on dataset C, it can be seen that even when confronted with unknown topological scenarios not seen in offline learning, the ASTGL model maintains a high level of \textit{Acc}, being more than 97\%. This reveals only a 2.5\% performance gap between known and unknown topological conditions. Hence, the ASTGL model can implement reliable online SVS monitoring under time-varying topological conditions.

\textit{2) Comparative Study}: To validate the SVS assessment performance of the proposed ASTGL model, this paper selects eight mainstream ML methods for comparative testing, namely DT, multilayer perceptron (MLP), CNN, transfer-learning–enhanced CNN (TL-CNN), LSTM, STGCN built upon an unweighted connectivity adjacency matrix (denoted by STGCN1), STGCN built upon an adjacency matrix represented by admittance (denoted by STGCN2) and transfer-learning-enhanced STGCN2 (TL-STGCN). For TL-CNN and TL-STGCN, transfer learning is performed by fine-tuning the pretrained models on a small subset of cases from dataset C, enabling better adaptation to unknown topological scenarios. Specifically, Table III shows the SVS assessment performance metrics for all models in known topological scenarios. To further validate the generalization capability of the ASTGL model, Table IV summarizes the test results of all models on dataset C. To ensure the fairness of the test results and reduce the workload and subjectivity of manual hyperparameter tuning, we utilized the Optuna Python library to optimize the hyperparameters of all methods. The detailed hyperparameter settings are provided in a downloadable supplementary file \cite{ref35} (see Table S1 in the file).

Firstly, for dataset A involving only one topological condition, all methods exhibit satisfactory SVS assessment performance. This is mainly because the stationarity of the single topological structure makes the SVS feature extraction and assessment process relatively simple, allowing models to efficiently learn and recognize fixed spatiotemporal SVS features. Secondly, when facing dataset B involving known topological change conditions, the accuracy of all models except ASTGL reduces significantly. Among them, DT and MLP are more severely affected, with their \textit{Acc} decreasing by 9.68\% and 6.89\%, respectively, compared to the ASTGL. This is mainly due to their limited learning capabilities, which prevent them from understanding the spatiotemporal features of transient responsive trajectories. The \textit{Acc} metrics of CNN and LSTM decrease by 3.57\% and 3.38\%, respectively, outperforming DT and MLP. This indicates that mining spatial and temporal features via DL networks can enhance the SVS assessment performance. The \textit{Acc} metric of TL-CNN improved compared to CNN, reaching 96.17\%. However, since they fail to comprehensively capture the temporal features behind transient responsive trajectories and the spatial features related to power grid structures, the SVS assessment performance of CNN and LSTM is inferior to that of GCN-based models. By capturing the spatial correlations between buses via the adjacency matrices, STGCN1 and STGCN2 can handle certain topological changes. However, they still fail to adaptively mine the dynamic spatiotemporal features under topological changes, with their SVS assessment accuracy on dataset B dropping to 97.73\% and 98.12\%, respectively. TL-STGCN achieves the \textit{Acc} metric of 98.64\%. This enhancement arises because fine-tuning on a small subset of samples from changed topologies helps mitigate feature mismatches between the training and testing network structures. In contrast, the \textit{Acc} metric of ASTGL steadily maintain above 99.2\%, with the other metrics remaining at exceptionally high levels as well. 
\begin{table}[ht]
	\centering
	\caption{SVS Assessment Performance Comparison Under Known Topological Changes}
	\label{tab4}
	% --- 减小列间距 ---
	\setlength{\tabcolsep}{3pt} 
	% --- 增加行间距 ---
	\renewcommand{\arraystretch}{1.0}
	
	% --- 使用 tabular* 并设置宽度为 \columnwidth ---
	\begin{tabular*}{\columnwidth}{l @{\extracolsep{\fill}} c c c c}
		\toprule
		\multirow{2}{*}[1pt]{\vspace{-7pt}\textbf{Model}} & \multicolumn{4}{c}{\textbf{Dataset A (One topology)}} \\
		\cmidrule(lr){2-5}
		& \textbf{\textit{Pre}/\% $\uparrow$} & \textbf{\textit{Rec}/\% $\uparrow$} & \textbf{\textit{F1}/\% $\uparrow$} & \textbf{\textit{Acc}/\% $\uparrow$} \\
		\midrule
		% --- Proposed 行 (无颜色) ---
		\textbf{Proposed} & \textbf{99.22} & \textbf{100.00} & \textbf{99.61} & \textbf{99.61} \\
		
		% --- 对比方法：应用 \diffcell 命令 ---
		TL-STGCN & \diffcell{0.51}{98.71} & \diffcell{0.39}{99.61} & \diffcell{0.45}{99.16} & \diffcell{0.46}{99.15} \\
		STGCN2 & \diffcell{0.64}{98.58} & \diffcell{0.13}{99.87} & \diffcell{0.39}{99.22} & \diffcell{0.39}{99.22} \\
		STGCN1 & \diffcell{0.40}{98.82} & \diffcell{1.44}{98.56} & \diffcell{0.92}{98.69} & \diffcell{0.92}{98.69} \\
		LSTM & \diffcell{3.41}{95.81} & \diffcell{1.44}{98.56} & \diffcell{2.44}{97.17} & \diffcell{2.48}{97.13} \\
		TL-CNN & \diffcell{2.54}{96.68} & \diffcell{1.17}{98.83} & \diffcell{1.87}{97.74} & \diffcell{1.90}{97.71} \\
		CNN & \diffcell{3.92}{95.30} & \diffcell{1.96}{98.04} & \diffcell{2.96}{96.65} & \diffcell{3.00}{96.61} \\
		MLP & \diffcell{6.44}{92.78} & \diffcell{1.04}{98.96} & \diffcell{3.84}{95.77} & \diffcell{3.98}{95.63} \\
		DT & \diffcell{4.60}{94.62} & \diffcell{5.87}{94.13} & \diffcell{5.24}{94.37} & \diffcell{5.22}{94.39} \\
		\midrule
		\multirow{2}{*}[1pt]{\vspace{-7pt}\textbf{Model}} & \multicolumn{4}{c}{\textbf{Dataset B (Topological changes)}} \\
		\cmidrule(lr){2-5}
		& \textbf{\textit{Pre}/\% $\uparrow$} & \textbf{\textit{Rec}/\% $\uparrow$} & \textbf{\textit{F1}/\% $\uparrow$} & \textbf{\textit{Acc}/\% $\uparrow$} \\
		\midrule
		% --- Proposed 行 (无颜色) ---
		\textbf{Proposed} & \textbf{98.84} & \textbf{99.61} & \textbf{99.22} & \textbf{99.22} \\
		
		% --- 对比方法：应用 \diffcell 命令 ---
		TL-STGCN & \diffcell{0.77}{98.07} & \diffcell{0.39}{99.22} & \diffcell{0.58}{98.64} & \diffcell{0.58}{98.64} \\
		% --- 这里是更正点 ---
		STGCN2 & \diffcell{1.16}{97.68} & \diffcell{1.04}{98.57} & \diffcell{1.10}{98.12} & \diffcell{1.10}{98.12} \\
		% --- 更正点结束 ---
		STGCN1 & \diffcell{0.93}{97.91} & \diffcell{2.08}{97.53} & \diffcell{1.50}{97.72} & \diffcell{1.49}{97.73} \\
		LSTM & \diffcell{3.36}{95.48} & \diffcell{3.38}{96.23} & \diffcell{3.37}{95.85} & \diffcell{3.38}{95.84} \\
		TL-CNN & \diffcell{1.76}{97.08} & \diffcell{4.42}{95.19} & \diffcell{3.09}{96.13} & \diffcell{3.05}{96.17} \\
		CNN & \diffcell{2.04}{96.80} & \diffcell{5.20}{94.41} & \diffcell{3.63}{95.59} & \diffcell{3.57}{95.65} \\
		MLP & \diffcell{6.95}{91.89} & \diffcell{6.76}{92.85} & \diffcell{6.85}{92.37} & \diffcell{6.89}{92.33} \\
		DT & \diffcell{7.76}{91.08} & \diffcell{11.96}{87.65} & \diffcell{9.89}{89.33} & \diffcell{9.68}{89.54} \\
		\bottomrule
		% --- 修改：使用 \end{tabular*} ---
\end{tabular*}
\begin{justify}
	\small \textbf{\textit{Note:}} Values in parentheses ($\% $) denote the performance drop relative to the proposed ASTGL model. The cell color intensity (green gradient) reflects this difference: deeper green indicates a larger performance gap.
\end{justify}
\end{table}
\begin{table}[ht]
	\centering
	\caption{SVS Assessment Performance Comparison Under Unknown Topological Changes}
	\label{tab5}
	\setlength{\tabcolsep}{3pt} 
	\renewcommand{\arraystretch}{1.0}
	\begin{tabular*}{\columnwidth}{p{1.45cm} @{\extracolsep{\fill}} c c c c}
		\toprule
		\multirow{2}{*}[1pt]{\vspace{-7pt}\textbf{Model}} & \multicolumn{4}{c}{\textbf{Dataset C (Topological changes)}} \\
		\cmidrule(lr){2-5}
		& \textbf{\textit{Pre}/\% $\uparrow$} & \textbf{\textit{Rec}/\% $\uparrow$} & \textbf{\textit{F1}/\% $\uparrow$} & \textbf{\textit{Acc}/\% $\uparrow$} \\
		\midrule
		\textbf{Proposed} & \textbf{98.16} & \textbf{97.14} & \textbf{97.65} & \textbf{97.66} \\
		TL-STGCN & \diffcell{1.54}{96.62} & \diffcell{0.39}{96.75} & \diffcell{0.96}{96.69} & \diffcell{0.97}{96.69} \\
		STGCN2 & \diffcell{2.85}{95.31} & \diffcell{1.95}{95.19} & \diffcell{2.40}{95.25} & \diffcell{2.40}{95.26} \\
		STGCN1 & \diffcell{2.46}{95.70} & \diffcell{4.42}{92.72} & \diffcell{3.46}{94.19} & \diffcell{3.38}{94.28} \\
		LSTM & \diffcell{1.78}{96.38} & \diffcell{7.15}{89.99} & \diffcell{4.58}{93.07} & \diffcell{4.35}{93.31} \\
		TL-CNN & \diffcell{3.23}{94.93} & \diffcell{4.55}{92.59} & \diffcell{3.90}{93.75} & \diffcell{3.83}{93.83} \\
		CNN & \diffcell{5.05}{93.11} & \diffcell{7.54}{89.60} & \diffcell{6.33}{91.32} & \diffcell{6.17}{91.49} \\
		MLP & \diffcell{8.20}{89.96} & \diffcell{5.07}{92.07} & \diffcell{6.65}{91.00} & \diffcell{6.76}{90.90} \\
		DT & \diffcell{9.49}{88.67} & \diffcell{13.65}{83.49} & \diffcell{11.65}{86.00} & \diffcell{11.24}{86.42} \\
		\bottomrule
		% --- 修改：使用 \end{tabular*} ---
\end{tabular*}
\end{table}

Thirdly, it can be seen from Table IV that under unknown topological change scenarios, ASTGL still exhibits the best performance, with its \textit{Acc} and \textit{F1} reaching 97.66\% and 97.65\%, respectively. In contrast, the performance of the eight comparative methods is severely degraded. Among them, DT and conventional DL methods such as MLP, CNN, LSTM and TL-CNN fail to adapt to changes in grid topological structures, resulting in poor performance under unknown topological conditions, with their \textit{Acc} falling below 95\%. Similarly, for STGCN1 and STGCN2, their \textit{Acc} indices also decreased by 3.38\% and 2.40\%, respectively. Through updating the pretrained model with a small number of new samples, TL-STGCN achieves an accuracy of 96.69\%. However, if the system topology changes frequently in practice, TL-STGCN would need to be repeatedly implemented, which may constrain its reliability and applicability in practical power grid operation scenarios. Conversely, ASTGL is able to outperform other methods in unknown topologies due to its unique module design. Through the joint effect of the AGLM and SAM modules, ASTGL can effectively adjusts the trainable weights of the graph representation matrix to adapt to topological variations. Furthermore, the residual structure of the Re-STGCN module enables deeper and more efficient system-wide SVS feature learning, thereby significantly enhancing its generalization performance.

Based upon the above result analyses, it can be concluded that the proposed ASTGL-enabled SVS assessment approach has higher reliability and thus better applicability in practical conditions with diverse unforeseen topological variations.
\begin{figure}[!t]
	\centering
	\includegraphics[width=3.5in]{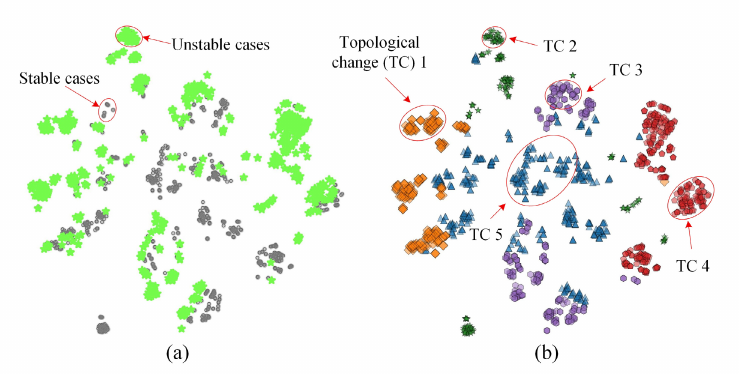}
	\caption{Visualization of case distribution in 2D t-SNE space. (a) distribution labeled by SVS status; (b) distribution labeled by topological changes.}
	\label{fig7}
\end{figure}

\textit{3) Visualization of Assessment Results}: To illustrate the structural generalization capability of the proposed ASTGL model, this paper uses the t-distributed stochastic neighbor embedding (t-SNE) \cite{ref36} to project high-dimensional features onto two dimensions: Fig. 8 visualizes case distribution of raw input features $\boldsymbol{X}$ from dataset C, labeled by SVS status $\boldsymbol{Y}$ and topological change (TC) groups, while Fig. 9 shows the learning results of the intermediate layer learning of the ASTGL model and several comparative methods. As shown in Fig. 8(a), the gray stable and green unstable cases form several overlapping clusters with no clear separating boundary. Fig. 8(b) further shows that samples within the same TC group tend to cluster, but notable overlap and unclear boundaries emerge between clusters of different TCs. Such inter-cluster ambiguity reflects that samples with the same SVS status but different topologies are mismatched in the input space.

From Fig. 9, it is observed that traditional DL models (CNN, LSTM) fail to achieve effective feature separation, as their gray stable and green unstable clusters remain heavily intertwined even at the final network layer L4. In contrast, TL-CNN, which is fine-tuned on a small subset of samples from the target domain (Dataset C), alleviated the feature mismatch problem caused by topological dissimilarities at the boundary of the CNN model. Next, for the GCN-based models, both STGCN1 and STGCN2 fail to form meaningful feature separation in the early layers L1 and L2, leading to a substantially larger number of FN samples-56 for STGCN1 and 37 for STGCN2. While TL-STGCN achieves reduced FN to 25 and FP to 26 through fine-tuning on a small subset of target-domain samples (Dataset C), such reliance on prior knowledge of unseen topologies is difficult for real-time SVS monitoring in operational power systems.
\begin{figure}[!t]
	\centering
	\includegraphics[width=3.5in]{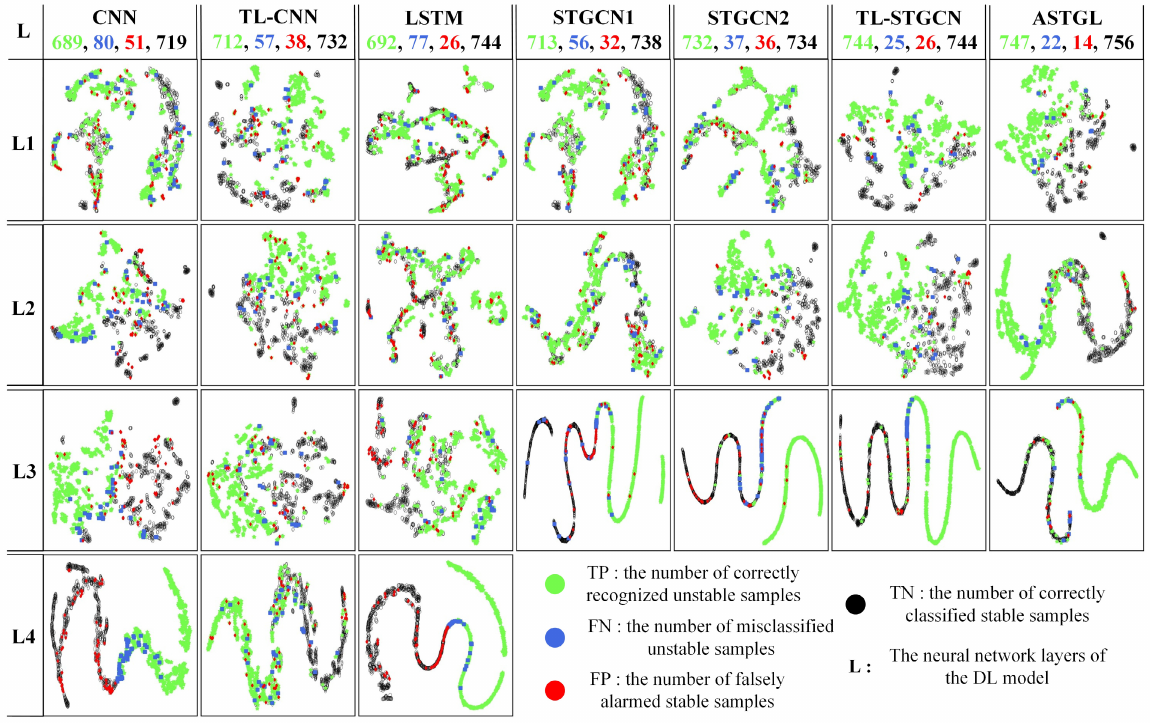}
	\caption{t-SNE enabled visualization of outputs of intermediate layers of different SVS assessment models on dataset C (with unknown topological changes).}
	\label{fig9} 
\end{figure}

In contrast to the above, the initial layers of ASTGL already show a certain degree of separation between different classes of samples. At layers L1 and L2, the integration of AGLM and SAM equips ASTGL with structure-aware learning capability under varying topologies. As a result, the subsequent Re-STGCN module can progressively align samples originating from different topological conditions within the embedding space. After passing through the CM at L3, the feature space becomes linearly separable, with stable and unstable cases clearly partitioned into two distinct regions. Importantly, the proposed ASTGL model achieves the minimum FP of 14 and FN of 22. The few remaining FN samples lie exactly at the boundary between the stable and unstable cases, indicating that the model has effectively driven nearly all high-risk unstable cases into the correct region. This implies that the ASTGL model has strong spatiotemporal feature mining capabilities under unknown topological scenarios, thereby enabling adaptive SVS assessment during online monitoring.

\textit{4) Cost and Efficiency:}
To comprehensively verify the assessment cost and computational efficiency of the proposed approach, this paper further examines it as well as other comparative methods in terms of four metrics. All the examinations are conducted on a server configured with an Intel i9-14900KF 6.00 GHz CPU, 64 GB RAM, and an NVIDIA GeForce RTX 4090 24 GB GPU. Among them, DT is realized by using the scikit-learn ML framework, while the other methods are realized by using the PyTorch DL framework. Table V summarizes the test results of all methods.

\begin{figure}[!t]
	\centering
	\includegraphics[width=3.5in]{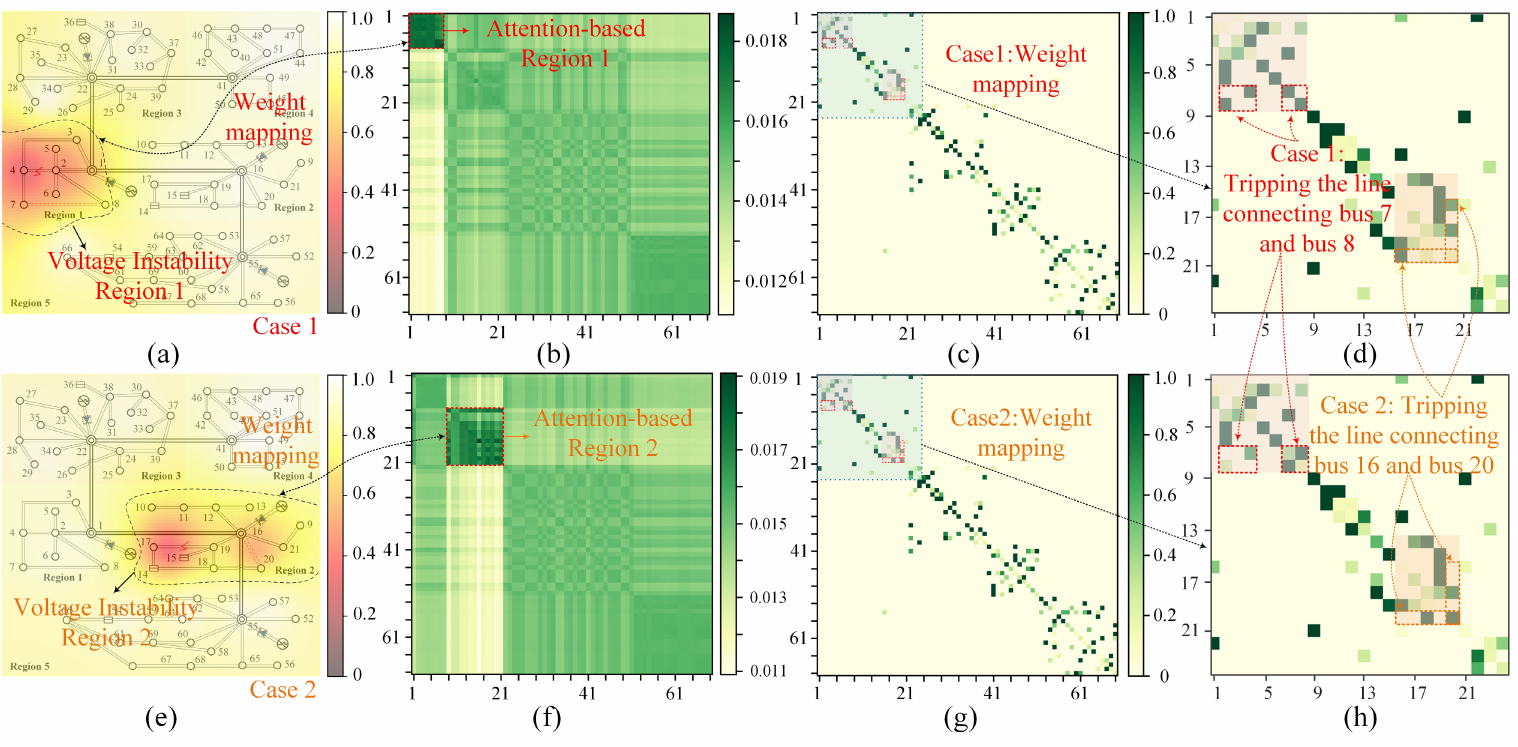} 
	\caption{Visualization of the spatial voltage distribution and graph matrix representation for two representative transient cases. (a) and (e) voltage distributions of Case 1 and Case 2 at t = 0.1 s; (b) and (f) attention-based grid representation matrix; (c) and (d) Case 1: admittance-based adjacency matrix and weight changes induced by topological variations; likewise, (g) and (h) show the corresponding results for Case 2.}
	\label{fig10}
\end{figure}
\begin{table}[!t]
	\centering
	\renewcommand{\arraystretch}{1.0} % 稍微比 1.0 大一点点，防止 booktabs 的横线贴住表头文字
	\scriptsize % 将整个表格的字体设置为 scriptsize
	\caption{Computational Efficiency Statistics of Different Methods}
	\label{tab6}
	\setlength{\tabcolsep}{3pt} % 进一步减小列间距
	\begin{tabular}{
			>{\raggedright\arraybackslash}p{0.14\linewidth} % Model (左对齐，宽度微调)
			>{\centering\arraybackslash}p{0.19\linewidth} % Offline training time/s (居中)
			>{\centering\arraybackslash}p{0.21\linewidth} % Offline computational memory/MB (居中)
			>{\centering\arraybackslash}p{0.16\linewidth} % Topology information dependence (居中)
			>{\centering\arraybackslash}p{0.18\linewidth} % Online SVS assessment/ms (居中)
		}
		\toprule
		\textbf{Model} & 
		\begin{tabular}[c]{@{}c@{}}\textbf{Offline} \\ \textbf{training} \\ \textbf{time/s}\end{tabular} & 
		\begin{tabular}[c]{@{}c@{}}\textbf{Offline} \\ \textbf{computational} \\ \textbf{memory/MB}\end{tabular} & 
		\begin{tabular}[c]{@{}c@{}}\textbf{Topology} \\ \textbf{information} \\ \textbf{dependence}\end{tabular} & 
		\begin{tabular}[c]{@{}c@{}}\textbf{Online SVS} \\ \textbf{assessment/ms}\end{tabular} \\ 
		\midrule % 替换了原来的 \hline
		
		DT & 984.5621 & 493.862 & No & 0.71 \\
		MLP & 1471.3873 & 1261.427 & No & 1.16 \\
		CNN & 3649.6185 & 1397.735 & No & 1.19 \\
		TL-CNN & 4897.2134 & 1568.442 & No & 1.21 \\
		LSTM & 5797.6572 & 1641.584 & No & 1.24 \\
		STGCN & 11756.5285 & 2074.163 & Yes & 1.48 \\
		TL-STGCN & 14893.7721 & 2716.928 & Yes & 1.59 \\
		Proposed & 13571.3744 & 2467.386 & No & 1.67 \\
		\bottomrule
	\end{tabular}
\end{table}
As can be seen, the offline training time and computational memory usage of the ASTGL model are 3.77 h and 2.41 GB, respectively. This is attributed to the relatively large number of hyperparameters in ASTGL, particularly those in the AGLM and Re-STGCN, which involve information passing between individual buses, thereby increasing the model's training burden. Yet it should be noted that, the offline training and computational cost of 3.77 h and 2.41 GB are affordable for ordinary servers and do not affect the online SVS application efficiency in practical power grids. In contrast, the STGCN model requires exact power topological information to ensure a certain level of SVS assessment performance, which introduces additional costs of topological information acquisition and is intractable in practical scenarios with highly frequent topological variations. Moreover, its transfer-learning extension, TL-STGCN, further increases offline computational demand because its additional trainable parameters significantly amplify both memory usage and training time. For online decision-making, ASTGL costs no more than 2 ms to issue SVS assessment results for a certain case. This means that the system can rapidly respond soon after fault clearance, thereby leaving sufficient time to help enhance system stability in online emergency situations.
\subsection{Adaptive Graph Analysis}
To investigate how the attention-based graph learning mechanism helps the whole approach gain desirable adaptability to topological changes, graph analysis is performed here via heatmap-based visualization. In particular, to conduct a comprehensive comparative test, the proposed attention-based grid representation matrix and an admittance-based adjacency matrix are selected for graph analysis. Without loss of generality, two typical cases are chosen for illustration (Case 1, tripping the line connecting bus 7 and bus 8 in Region 1; Case 2, tripping the line connecting bus 16 and bus 20 in Region 2). For each case, the spatial voltage distribution at a specific transient time instant is visualized, and the differences between the above-mentioned two graph matrices are visually compared to analyze which matrix can better capture the networked spatial correlations within different cases.

For Case 1, Fig. 10(a) clearly shows that after the fault on the line connecting bus 2 and bus 4, the voltage magnitudes of all buses in region 1 drop below 0.8 pu. However, the admittance-based matrix in Fig. 10(c)-(d) only increases the weights associated with buses 2, 4, 7, and 8, while assigning limited attention to other buses such as buses 1, 3, 5, and 6, despite these locations also experiencing severe voltage sags. Similarly, in Case 2, the admittance matrix shown in Fig. 10(g) and (h) also fails to focus on the severely voltage-saggy buses in system region 2, such as bus 16 and bus 18. This indicates that such a pattern of weight allocation fails to capture the transient voltage distribution from a system-wide perspective. 

In contrast, as shown in Fig. 10(b) and (f), the attention-based grid matrix can dynamically characterize the variations of spatial correlations between individual buses in different transient cases. For instance, in Case 1, the learned matrix effectively characterizes the spatial correlations within region 1 while also capturing the interactions between this region and other parts of the grid. In Case 2, it assigns higher weights to bus 16 and bus 20, which are involved in tripping the connecting line. This reveals that the attention-based matrix contributes to automatic perception of system topological changes without any prior knowledge. Besides, it suggests that the matrix’s element values can indicate local voltage sag regions to some extent, which may be helpful for potential emergency control schemes to select critical control sites. Owing to these merits, the ASTGL approach is thus able to more adaptively capture the variable spatial correlations than existing alternatives built upon other fixed adjacency matrices.
\begin{table}[!t]
	\centering
	\renewcommand{\arraystretch}{1.0}
	\caption{Statistics of Generated Datasets}
	\label{tab7}
	\resizebox{\linewidth}{!}{
		\begin{tabular}{>{\centering\arraybackslash}m{3.5cm} >{\centering\arraybackslash}m{1cm} >{\centering\arraybackslash}m{1.5cm} >{\centering\arraybackslash}m{1.5cm} >{\centering\arraybackslash}m{1.2cm}}
			\toprule
			\multirow{2}{=}{\centering Dataset} & \multirow{2}{=}{\centering Split} & \multicolumn{2}{c}{Labels} & \multirow{2}{=}{\centering Total} \\ \cmidrule{3-4}
			& & Stable & Unstable & \\ \midrule
			
			% Dataset D (Known topologies)
			\multirow{2}{=}{\centering D (Topological changes)} & Train & 4070 & 4068 & 8138 \\
			& Test & 1018 & 1017 & 2035 \\ \midrule
			
			% Dataset E (Unknown topologies)
			E (Topological changes) & Test & 1018 & 1017 & 2035 \\
			\bottomrule
		\end{tabular}
	}
\end{table}
\begin{table}[ht]
	\centering
	\caption{SVS Assessment Performance Comparison Under Topological Changes}
	\label{tab8}
	% --- 减小列间距 ---
	\setlength{\tabcolsep}{3pt} 
	% --- 增加行间距 ---
	\renewcommand{\arraystretch}{1.0}
	
	% --- 使用 tabular* 并设置宽度为 \columnwidth ---
	\begin{tabular*}{\columnwidth}{p{1.44cm} @{\extracolsep{\fill}} c c c c}
		\toprule
		\multirow{2}{*}[1pt]{\vspace{-7pt}\textbf{Model}} & \multicolumn{4}{c}{\textbf{Dataset D (Known topological changes)}} \\
		\cmidrule(lr){2-5}
		& \textbf{\textit{Pre}/\% $\uparrow$} & \textbf{\textit{Rec}/\% $\uparrow$} & \textbf{\textit{F1}/\% $\uparrow$} & \textbf{\textit{Acc}/\% $\uparrow$} \\
		\midrule
		% --- Proposed 行 (无颜色) ---
		\textbf{Proposed} & \textbf{99.11} & \textbf{99.02} & \textbf{99.07} & \textbf{99.07} \\
		
		% --- 对比方法：应用 \diffcell 命令 ---
		TL-STGCN & \diffcell{1.36}{97.75} & \diffcell{0.99}{98.03} & \diffcell{1.18}{97.89} & \diffcell{1.18}{97.89} \\
		
		STGCN2 & \diffcell{2.82}{96.29} & \diffcell{2.07}{96.95} & \diffcell{2.45}{96.62} & \diffcell{2.46}{96.61} \\
		
		STGCN1 & \diffcell{4.08}{95.03} & \diffcell{3.05}{95.97} & \diffcell{3.57}{95.50} & \diffcell{3.59}{95.48} \\
		
		LSTM & \diffcell{6.44}{92.67} & \diffcell{4.53}{94.49} & \diffcell{5.50}{93.57} & \diffcell{5.56}{93.51} \\
		
		TL-CNN & \diffcell{3.37}{95.74} & \diffcell{4.03}{94.99} & \diffcell{3.71}{95.36} & \diffcell{3.69}{95.38} \\
		
		CNN & \diffcell{5.23}{93.88} & \diffcell{6.98}{92.04} & \diffcell{6.12}{92.95} & \diffcell{6.05}{93.02} \\
		
		MLP & \diffcell{8.53}{90.58} & \diffcell{12.00}{87.02} & \diffcell{10.30}{88.77} & \diffcell{10.08}{88.99} \\
		
		DT & \diffcell{14.45}{84.66} & \diffcell{11.11}{87.91} & \diffcell{12.82}{86.25} & \diffcell{13.07}{86.00} \\
		
		\midrule
		\multirow{2}{*}[1pt]{\vspace{-7pt}\textbf{Model}} & \multicolumn{4}{c}{\textbf{Dataset E (Unknown topological changes)}} \\
		\cmidrule(lr){2-5}
		& \textbf{\textit{Pre}/\% $\uparrow$} & \textbf{\textit{Rec}/\% $\uparrow$} & \textbf{\textit{F1}/\% $\uparrow$} & \textbf{\textit{Acc}/\% $\uparrow$} \\
		\midrule
		% --- Proposed 行 (无颜色) ---
		\textbf{Proposed} & \textbf{96.98} & \textbf{98.03} & \textbf{97.51} & \textbf{97.49} \\
		
		% --- 对比方法：应用 \diffcell 命令 ---
		TL-STGCN & \diffcell{2.77}{94.21} & \diffcell{1.96}{96.07} & \diffcell{2.38}{95.13} & \diffcell{2.40}{95.09} \\
		% --- 这里是更正点 ---
		STGCN2 & \diffcell{3.42}{93.56} & \diffcell{5.11}{92.92} & \diffcell{4.27}{93.24} & \diffcell{4.22}{93.27} \\
		% --- 更正点结束 ---
		STGCN1 & \diffcell{5.92}{91.06} & \diffcell{5.90}{92.13} & \diffcell{5.92}{91.59} & \diffcell{5.94}{91.55} \\
		
		LSTM & \diffcell{8.48}{88.50} & \diffcell{11.01}{87.02} & \diffcell{9.76}{87.75} & \diffcell{9.69}{87.80} \\
		
		TL-CNN & \diffcell{7.09}{89.89} & \diffcell{7.08}{90.95} & \diffcell{7.09}{90.42} & \diffcell{7.12}{90.37} \\
		
		CNN & \diffcell{12.45}{84.53} & \diffcell{12.09}{85.94} & \diffcell{12.28}{85.23} & \diffcell{12.38}{85.11} \\
		
		MLP & \diffcell{14.33}{82.65} & \diffcell{17.01}{81.02} & \diffcell{15.68}{81.83} & \diffcell{15.48}{82.01} \\
		
		DT & \diffcell{19.46}{77.52} & \diffcell{18.68}{79.35} & \diffcell{19.08}{78.43} & \diffcell{19.31}{78.18} \\
		\bottomrule
		% --- 修改：使用 \end{tabular*} ---
\end{tabular*} 
\end{table}
\subsection{Verification of Scalability in Large-Scale Power Systems}
In order to verify the scalability of the proposed method on a large-scale power system, an extended case study is conducted on sub-system 2, as illustrated in Fig. 4. The case generation procedures remain consistent with those used in Section V-A. Accordingly, fifteen representative 220-kV transmission lines are chosen, and one or two of these lines are randomly tripped to generate \(N\)-1 and \(N\)-2 topological variations. Based on these variations and batch-TD simulations, the generated datasets are statistically summarized in Table VI.

For dataset D involving known topological change conditions, Table VII shows that as the system scale and the complexity of topological variations increase, the performance stability of all comparative models generally declines. The STGCN1 and STGCN2 exhibit pronounced performance degradation, as their static adjacency structures fail to capture the increasingly diverse and complex spatial dependencies in the enlarged network. Although TL-STGCN benefits from fine-tuning and achieves moderate improvement, its accuracy still decreases to 97.89\%. In contrast, the proposed ASTGL delivers the highest and most stable performance (Acc 99.07\%, F1 99.07\%). Furthermore, when facing dataset E involving unknown topological change conditions, the performance gap between ASTGL and the comparative models becomes even more pronounced. ASTGL maintains an accuracy of 97.49\%, outperforming TL-STGCN. In comparison, the fixed-graph models degrade more significantly, with STGCN1 and STGCN2 dropping to 91.55\% and 93.27\%, respectively. Overall, the key module design of ASTGL enables the model to autonomously perceive and adapt to previously unseen topological structures in large-scale systems, thereby demonstrating superior scalability and enhanced practical applicability for real-world power systems.
\section{Conclusion}
This study proposes an intelligent ASTGL approach for SVS assessment against time-varying topological conditions. By integrating an adaptive graph learning network with a spatial attention mechanism, the ASTGL approach effectively captures the inherent spatial correlations within a specific grid under diverse topological changes. This desirable adaptability enables the ASTGL approach reliably to achieve structural change-aware SVS assessment in practical contexts. Extensive numerical tests on two representative sub-systems of a realistic provincial power grid in South China sufficiently verify the superiority of the proposed approach. In particular, the ASTGL approach achieves an SVS assessment accuracy of 97.66\% on Sub-system 1 and 97.49\% on Sub-system 2 under diverse scenarios involving topological variations, demonstrating its strong scalability and adaptability to large-scale and structurally dynamic power networks. Compared with other learning schemes, it has no reliance on knowledge of exact system topological information and adapts better to unknown topological scenarios, being more applicable in practical systems involving frequently changing topologies.

In future research, physical mechanisms and laws w.r.t. the SVS issue could be incorporated into the ASTGL framework to guide the learning procedure to better mine the inherent SVS characteristics for SVS assessment performance enhancement.
%\colorbib{ref17}{blue}
%\colorbib{ref18}{blue}
%\colorbib{ref19}{blue}
%\colorbib{ref20}{blue}
%\colorbib{ref21}{blue}
%\colorbib{ref22}{blue}
%\colorbib{ref23}{blue}
%\colorbib{ref35}{blue}
\bibliography{reference} % 指定 .bib 文件名（无需后缀）
\end{document}